\documentclass[twocolumn,final]{svjour3}

\smartqed 
\usepackage{graphicx}
\usepackage{natbib}

\newcommand{\Lsun}{\hbox{$L_\odot$}}
\newcommand{\Msun}{\hbox{$M_\odot$}}
\newcommand{\UCHII}{\hbox{UCH{\sc ii}}}

\journalname{Astrophysics and Space Science}

\begin{document}

\title{APEX survey of southern high mass star forming regions
}

\author{C. Hieret \and S. Leurini \and K. M. Menten \and P. Schilke \and S. Thorwirth \and F.~Wyrowski}

\institute{Carolin Hieret \at
              MPIfR, Auf dem H\"ugel 69, 53121 Bonn, Germany\\
	      Tel.: +49(228) 525 491\\
	      Fax : +49(228) 525 435
              \\
              \email{chieret@mpifr-bonn.mpg.de}
}

\date{Received: date / Accepted: date}

\maketitle

\begin{abstract}
A systematic study of a large sample of sour-ces, covering a wide range in galactocentric distances, masses and luminosities, is a fast and efficient way of obtaining a good overview of the different stages of high-mass star formation.
With these goals in mind, we have started a survey of 40 color selected IRAS sources south of $-$20$^{\circ}$ declination with the APEX telescope on Chajnantor, Chile.
  Our first APEX results already demonstrate that the selection criteria were successful, since some of the sources are very rich in molecular lines. 

\keywords{Astrochemistry \and Surveys \and Stars: Formation \and ISM: Molecules \and Submillimeter }

\end{abstract}

\section{Introduction}
\label{intro}

The field of high mass star formation has seen a lot of progress in recent years.
The main theoretical problem of radiation pressure threatening to stall accretion has been adressed with theories such as increased turbulent accretion \citep{2003ApJ...585..850M}, non-spherical accretion through disks \citep{2002ApJ...569..846Y} but also coalescence \citep{2005AJ....129.2281B} and competitive accretion \citep{2004MNRAS.349..735B}.
From an observational point of view, the study of high mass star formation is taking a huge step forward with the availability of new submm instruments such as APEX and the SMA. Despite the ongoing effords to study the earliest stages of high mass star formation \citep{2002ApJ...566..945B,2004A&A...426...97F,2005MNRAS.363..405H} no survey so far contained a large number of sources observed in the whole submm regime, including the 350~$\mu$m window. \\
Therefore we have started a survey of 40 high mass star forming regions, based on color selected IRAS point source criteria, which we are observing between 230~GHz and 950~GHz.
The spectral lines in the submm regime (see Table \ref{tab:1}) let us probe higher levels of excitation and densities and are therefore more appropriate for the study of high mass star forming regions than mm studies alone.
Our selection criteria are such that this large sample will contain sources at very different stages of evolution and will allow a consistent and comparable analysis of these regions.

\section{Observations}
\label{obs}
The data were taken mainly with the Atacama Path-finder Experiment (APEX), a 12~m submillimeter telescope on Cerro Chajnantor, Chile at 5100~m \citep{2006A&A...454L..13G}.
We used both the APEX2a heterodyne receiver \citep{2006A&A...454L..13G} and the dual frequency 460/810 GHz Flash heterodyne receiver \citep{2006A&A...454L..13G} for our observations.
The APEX2a receiver has a beamwidth of 18'' at 345 GHz and the FLASH beam-widths are 14'' at 460 GHz and 7'' at 810 GHz. 
The pointing was accurate within 2''.\\

\subsection{Frequency setups}
\label{freq:setup}

To study the physical, chemical and kinematical properties of the sources, we are observing them  in a range of frequency setups (see Table \ref{tab:1}).
We have chosen to study the density and temperatures of the sources using H$_2$CO - and CH$_3$OH bands. Both, being slightly asymmetric rotors, are useful tracers of temperature and densities \citep{2004A&A...422..573L,1993ApJS...89..123M}, and have a multitude of transitions in the submm range. Furthermore they can be found in many star forming regions and trace gas over a wide density range.\\
The CO transitions will be used to study the extended envelopes of the sources, including outflows, whereas sources with a hot core behaviour are observed further in transitions of HCN, to examine the densest gas.
All sources are also observed in C$^{17}$O to derive the total column density of the molecular gas.

\begin{table}

\caption{Frequency Setups}
\label{tab:1}      
\begin{tabular}{|l|c|c|}
\hline\noalign{\smallskip}
molecule & frequency  &  tracer  \\
\noalign{\smallskip}\hline\noalign{\smallskip}
CH$_3$OH(6-5) & 290 GHz  & density \& temperature \\
CH$_3$OH(7-6) & 338 GHz & density \& temperature \\
H$_2$CO(4-3) & 291 GHz & density \& temperature \\
H$_2$CO(5-4)  & 363 GHz & density \& temperature \\
H$_2$CO(6-5)  & 437 GHz & density \& temperature \\
CO(3-2) & 345 GHz & outflow tracer \\
CO(4-3) & 461 GHz & outflow tracer \\
CO(7-6) & 806 GHz & outflow tracer \\
C$^{17}$O(3-2) & 338 GHz & molecular column density \\
HCN(4-3) & 354 GHz & dense gas  \\
HCN(9-8) & 797 GHz & dense gas  \\
HCO$^+$(4-3) & 357 GHz & dense gas, outflows\\
\noalign{\smallskip}\hline
\end{tabular}
\end{table}

\section{The source sample}
\label{sample}

\begin{figure}

  \includegraphics[angle=-90,width=8cm]{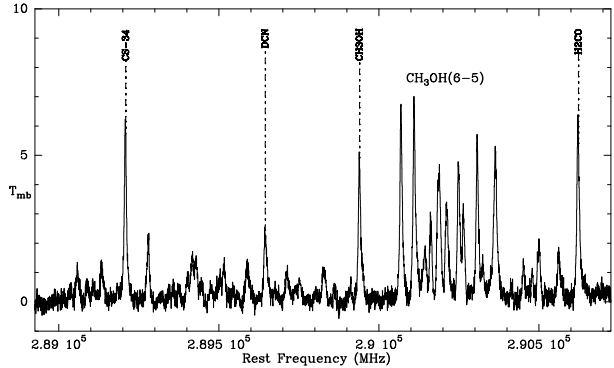}

\caption{ The CH$_3$OH(6-5) band in IRAS 17233-3606. This source turned out to be extremely rich in molecular lines. In this graph, only a sample of the identified lines is shown.}
\label{fig:1}      
\end{figure}

The sample contains 40 IRAS sources based on the samples of \citet{2004A&A...426...97F} and \citet{1998MNRAS.301..640W}, including sources with and without
radio continuum and/or maser associations (see Table \ref{tab:2}).\\ The inital selection of the
IRAS sources was done following the color selection criteria from \citet{1989ApJ...340..265W}, which select objects with envelope characteristics similar to cloud cores with embedded \UCHII\/ regions. In those sources, the existence of cm emission pinpoints to an advanced and more massive stage of star formation, whereas CH$_3$OH masers are seen as signs of the earliest stages of star formation.\\
IR luminosities range from 10$^3$ to a few 10$^5$ \Lsun\/ in our sample and with masses from 100 to 10$^5$
\Msun\/.  They were selected to encompass a broad range
of evolutionary stages, since the aim of this study is to classify
properties of high mass star forming regions on a statistically
significant sample.  Our preliminary analysis shows that
the sources have very diverse properties and range from very chemically rich hot cores
(see Fig. \ref{fig:1} for a spectrum of 17233-3606) to sources that contain barely CH$_3$OH and little
more.

\begin{table}
\begin{center}

\caption{Sample Statistics, showing number of sources with and without cm emission and maser (CH$_3$OH, H$_2$O and OH) emission.}
\label{tab:2}      
\begin{tabular}{|l|c|c|}
\hline\noalign{\smallskip}
&\multicolumn{2}{c|}{masers}\\
\cline{2-3}
cm emission & yes & no\\

\cline{1-1}\cline{2-3}
yes & 20 & 4\\
no &  11 & 5 \\
\noalign{\smallskip}\hline
\end{tabular}
\end{center}
\end{table}

\section{Work in progress}
\label{wip}

\begin{figure}

  \includegraphics[angle=-90,width=8cm]{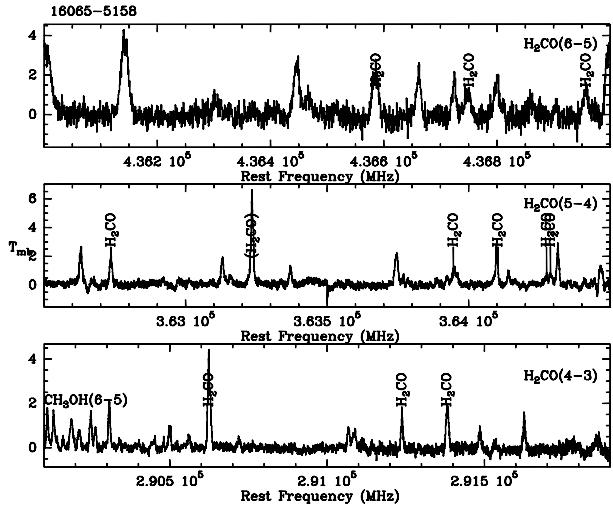}

\caption{H$_2$CO bands in IRAS 16065-5158 at 290 GHz, 363 GHz and 436 GHz. For the excitation analysis of this molecule, a LVG code (Leurini et al., 2004) will be used to model this large range of excitation conditions.}
\label{fig:2}      
\end{figure}

The excitation analysis is partly done in LTE with the XCLASS code \citep{1999pcim.conf..330S,2005ApJS..156..127C} and with an LVG code \citep{2004A&A...422..573L} for CH$_3$OH(6-5), CH$_3$OH(7-6) and the H$_2$CO bands. 
Both molecules are good tracers of density and temperature and have the advantage of having a large number of transitions in the observed frequency setups (see Fig.~\ref{fig:2} for the spectra of H$_2$CO bands in 16065-5158 between 290 GHz and 436 GHz). With these spectra we have access to levels between a few 10 to over 300~K above the ground state.   \\

\begin{figure}

  \includegraphics[angle=-90,width=8cm]{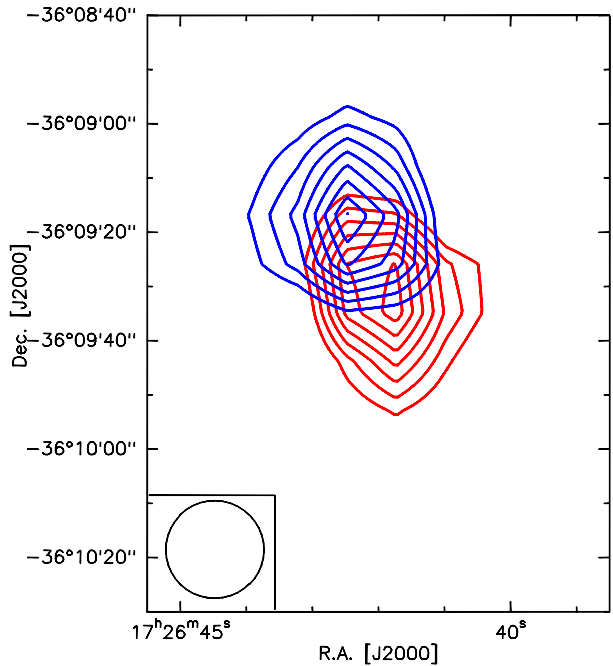}

\caption{Outflow of IRAS 17233-3606. This CO(3-2) map was observed with APEX.}
\label{fig:3}     
\end{figure}

Apart from modeling densities and temperatures, the data are used for kinematical analysis of the line shape, suggesting outflows which we intend to map (see Fig. \ref{fig:3}), and for possible signatures of infall, which can be seen in the HCO$^+$ data.
In addition to providing the physical and kinematical properties, our data also allow us to obtain a chemical inventory of the sources.\\ 
We are using GLIMPSE data from the Spitzer Archive to study the large scale environment of the sources at infrared wavelengths. Over the course of 2007, this will be complemented with large submm maps from the LABOCA and SABOCA bolometers at APEX and high frequency maps from the CHAMP+ heterodyne array.\\
To obtain a better idea of the diverse conditions in high mass star forming regions, we have just started to extend our survey to sources in the outer Galaxy.
Follow-up studies have already been started with ATCA (3~mm observations in selected sources) and the SMA (imaging the bipolar outflow in 17233-3606, Fig. \ref{fig:3}).
These interferometer observations with their high spatial resolution allow us to study in detail the structure of the star forming cores and their association with masers and \UCHII\/ regions in their surroundings.

\section{Conclusions}

This large sample of sources will allow us to study and specify the evolutionary stages of high mass star formation on a statistically significant sample. 
In addition, by learning more about the high mass star forming regions of the southern hemisphere, which have been little studied so far at submm wavelengths, we will create a valuable sample of sources for follow-up studies with ALMA.

\begin{acknowledgements}
C. Hieret is a fellow of the Studienstiftung des deutschen Volkes and a member of the International Max Planck Research School for Radio and Infrared Astronomy. 
\end{acknowledgements}

\bibliographystyle{/aux/pc20005a/chieret/hot_cores/database/lit/aa}

\bibliography{/aux/pc20005a/chieret/hot_cores/database/lit/prom}

\end{document}